\documentclass[conference]{IEEEtran}
\usepackage{color}
\usepackage{multicol}
\usepackage{graphicx}
\usepackage{multirow}
\usepackage{booktabs}
\usepackage[table]{xcolor}
\usepackage{algorithm}
\usepackage{algorithmic}
\usepackage{amsmath,amssymb,amsthm}
\usepackage{epic,multibox,fancybox}
\usepackage{float}
\usepackage{blindtext}

\usepackage{subfigure}
\usepackage{graphicx}
\usepackage{caption}
\ifCLASSINFOpdf

\else

\fi
\begin{document}

\title{M-GEAR: Gateway-Based Energy-Aware Multi-Hop Routing Protocol for WSNs}

\author{Q. Nadeem$^{1}$, M. B. Rasheed$^{1}$, N. Javaid$^{1}$, Z. A. Khan$^{2}$, Y. Maqsood$^{2}$, A. Din$^{2}$\\
        $^{1}$COMSATS Institute of Information Technology, Islamabad, Pakistan.\\
        $^{2}$Faculty of Engineering, Dalhaousie University, Halifax, Canada.\\
        $^{3}$Abasyn University, Peshawar, Pakistan.\\

     }

\maketitle
\begin{abstract}
In this research work, we advise gateway based energy-efficient routing protocol (M-GEAR) for Wireless Sensor Networks (WSNs). We divide the sensor nodes into four logical regions on the basis of their location in the sensing field. We install Base Station (BS) out of the sensing area and a gateway node at the centre of the sensing area. If the distance of a sensor node from BS or gateway is less than predefined distance threshold, the node uses direct communication. We divide the rest of nodes into two equal regions whose distance is beyond the threshold distance. We select cluster heads (CHs)in each region which are independent of the other region. These CHs are selected on the basis of a probability. We compare performance of our protocol with LEACH (Low Energy Adaptive Clustering Hierarchy). Performance analysis and compared statistic results show that our proposed protocol perform well in terms of energy consumption and network lifetime.

\end{abstract}
\IEEEpeerreviewmaketitle
\section{Introduction}
A key concern in WSN technology is to enhance the network lifetime and to reduce the energy consumption of the sensor network. Wireless sensor nodes are dispersed typically in sensing area to monitor earthquake, battle field, industrial environment, habitant monitoring~\cite{mainwaring2002wireless}, agriculture field~\cite{burrell2004vineyard}, physical atmosphere conditions and smart homes. Sensor nodes sense the environment, gather information and transmit to BS through wireless link.\\
\indent Due to escalating in Micro-Electro-Mechanical System technology, now it is possible to set up thousands or millions of sensor nodes. The intense deployment of WSN makes it quite difficult to recharge node batteries. Therefore, a key subject for WSNs is to curtail power expenditure of sensor nodes to prolong network lifetime. Many clustering based algorithms~\cite{ye2005eecs}~\cite{li2005energy} are proposed.
Clustering is a technique in which network energy consumption is well managed by minimizing the transmission range of the sensors. In this modus operandi, CH manages the group communication with the BS. Sensor nodes no longer transmit data directly to the BS instead CHs receive the whole group messages, aggregates and forwards to the BS.\\
\indent All nodes in cluster transmit their data to corresponding CH. The CH issues a Time Division Multiple Access (TDMA) schedule for its member nodes to avoid collision. Each member node transmits its data to CH only in defined allocated time slot therefore, sensor nodes turn off their transceivers otherwise. TDMA scheduling encourages saving energy of sensor nodes and these nodes stay alive for longer period. As a rule, each member node transmits its data to nearby CH therefore; sensor nodes require minimum energy for data transmission. CHs perform computation on collected data and filter out the redundant bits, it reduces the amount of data that has to forward to the BS. Consequently, transmission energy of sensors reduce to significant amount.
In this research work, we design a gateway based energy-aware multi-hop routing protocol.\\
\indent The impulse behind this work is to trim down the energy consumption of sensor nodes by logically dividing the network into four regions. We use different communication hierarchy in different regions. Nodes in one region communicate directly to BS while nodes in region 2 communicate directly to gateway node. Nodes in other two regions use clustering hierarchy and sensor nodes transmit their data to gateway node through their CHs. Gateway node assists in defining clusters and issues a TDMA schedule for CHs. Each CH issues its own TDMA schedule for its member nodes.\\
The rest of the paper is ordered as follows: section 2 briefly review the related work. In section 3, we describe motivation for this work. Section 4 describes the network model. Proposed algorithm is explained in section 5. In section 6, we  define the performance parameters and show the performance of our proposed protocol by simulations and compare it with LEACH. Finally, section 7 gives conclusion.
\section{Related work}
Energy consumption and network lifetime are the most important features in the design of the wireless sensor network. This study present clustering based routing for WSNs. Many clustering based protocols are homogeneous, such as LEACH~\cite{heinzelman2000energy} PEGASIS~\cite{lindsey2002pegasis} and HEED~\cite{younis2004heed}.
CHs collect data from its members or slave nodes, aggregate and than forward to faraway located BS. This process overloads the CH and it consumes lot of energy. In LEACH, the CHs are selected periodically and consume uniform energy by selecting a new CH in each round. A node become CH in current round on the basis of probability p. LEACH performs well in homogenous network however, this protocol is not considered good for heterogeneous networks as shown in~\cite{smaragdakis2004sep}.\\
\indent In~\cite{loscri2005two} author presented another clustering protocol(TL-LEACH). This protocol describes two level clustering scheme which performs well in terms of minimum energy consumption of network. There are two levels of CHs, level one CHs and level two CHs. Level one CHs connect with their corresponding member sensor nodes. CHs at second level create clusters from CHs of level one.
TL-LEACH scheme is potentially more dispense therefore; the load of the network on the sensors is well shared which results in long lived sensor network.\\
\indent In PEGASIS~\cite{lindsey2002pegasis} nodes form a chain to transfer data from source to sink. In chain formation process each node connect with next node. The chain formation process require global knowledge of sensor nodes, hence, it is very difficult to implement this topology.\\
\indent Another clustering based protocol is HEED in which CHs are selected on the base of a probability. The probability of a node to become CH is related to the residual energy of the node. In HEED, it is possible that the nodes with minimum residual energy acquire larger probability to become CH.\\
\indent A PEGASIS based mobile sink scheme is proposed in~\cite{jafri2013maximizing}. The sink moves along its trajectory and stays for a sojourn time at sojourn location to guarantee complete data collection. A similar sink mobile based technique is proposed in~\cite{akbar2013modeling}.
\indent SEP protocol is designed for heterogeneous nodes. Nodes in SEP are heterogenous in terms of their initial energy, called normal nodes and advance nodes. The probability to become CH depends on the initial energy of the node. Performance of SEP in multi level Heterogeneous networks is not good.\\
\indent An Energy Efficient Unequal Clustering (EEUC) protocol is presented which tries to balance the energy consumption of the network. EEUC divide the network field into unequal clusters. In EEUC, there are some nodes in network that are not associated with any cluster, therefore, they are isolated inside the network.\\
\indent On adaptive energy-efficient scheme for transmission (EAST) is proposed in~\cite{tahir2013adaptive}. This scheme use open-looping feedback process for temperature-aware link quality estimation, whereas closed-loop feedback process divides network into three logical regions to minimize overhead of control packets. In~\cite{manzoor2013q} Quadrature-LEACH (Q-LEACH) for homogenous networks is proposed. This scheme maximize the throughput, lifetime of network and stability period of the network.\\
\indent Latif et al.~\cite{latif2013divide} presented Divide-and-Rule (DR) scheme. DR technique used for static clustering also for the selection of CH. This scheme avoids probabilistic selection of CH instead it elects fixed number of CH. Away Cluster Head (ACH) prtocol for WSN is proposed in~\cite{javaid2013ach}. This protocol efficiently maximize the stability period and throughput. J. Kulik et al.~\cite{heinzelman1999adaptive} proposed sensor Protocols for Information Via Negotiation (SPIN). In SPIN, a node advertise its sensed data to its neighbors about the kind of the data it sensed. An interested neighboring node will send a request for a copy of data to originating node. In this way, the entire nodes in the network acquire this data. The drawback of this approach is that, there is no guarantee of data delivery to each node in the network because if the node is interested in data from distant source node then data will not deliver to interested node. This protocol is not suited for applications where reliable data delivery priority is on top.\\
A hybrid protocol Hybrid Energy Efficient Reactive Protocol for WSN is proposed in~\cite{javaid2013heer}. In this protocol, CH is selected based on the residual energy of node and average energy of network.
\section{Motivation}

Due to the fact that clustering protocols consume less energy, these protocols for WSNs have gained extensive acceptance in many applications. Many state of the art WSN protocols exploit cluster based scheme at manifold levels to minimize energy expenditures. CH in most cluster based protocols is selected on the base of  probability. It is not obvious that CHs are distributed uniformly throughout the sensor field. Therefore, it is quite possible that the selected CHs concentrate in one region of the network. Hence, a number of nodes will not get any CHs in their environs.
Similarly some protocols used unequal clustering and try to use recourses proficiently.\\
Multiple level clustering hierarchy has following major drawbacks.\\
\begin{itemize}
 \item In multiple level schemes, one CH forward data to other CH which relays data to BS. If relay CH is faraway, than it is necessary for forwarder CH to transmit data with high power.\\
 \item In clustering protocols, a member node decides itself whether to become CH or not. It is possible that some distant nodes are selected as CHs. Therefore, these nodes consume lot of energy to forward data to BS. Hence, these nodes will die early\\
     \end{itemize}
In this article, our goal is to design a gateway based energy aware multi-hop routing protocol. This approach meets the following points.\\
\begin{itemize}
\item Network is divided into regions and aid of gateway node reduces the average transmission distance. Hence, it saves network energy and prolong network lifetime.\\
\item CH selection in each region is independent of other regions so, there is definitely a CH exist in each region.\\
    \end{itemize}

\section{Network Model}
In this article, we assume ${S}$ sensors which are deployed randomly in a field to monitor environment. We represent the \emph{i-th} sensor by s$_{i}$ and consequent sensor node set S= {{s$_{1}$, s$_{2}$,....., s$_{n}$ }}.
We assume the network model shown in fig 1.\\
\begin{itemize}
\item We deploy the BS faraway from the sensing field. Sensor nodes and the BS are
  stationary after deployment.
\item A gateway node is deployed in the same network field at the centre of the network.
\item Gateway node is stationary after deployment and rechargeable.
\item We use  homogeneous sensor nodes with same computational and sensing capabilities.
\item Each sensor node is assigned with a distinctive identifier (ID).
\end{itemize}
\begin{figure}
\begin{center}
\includegraphics[scale=0.2]{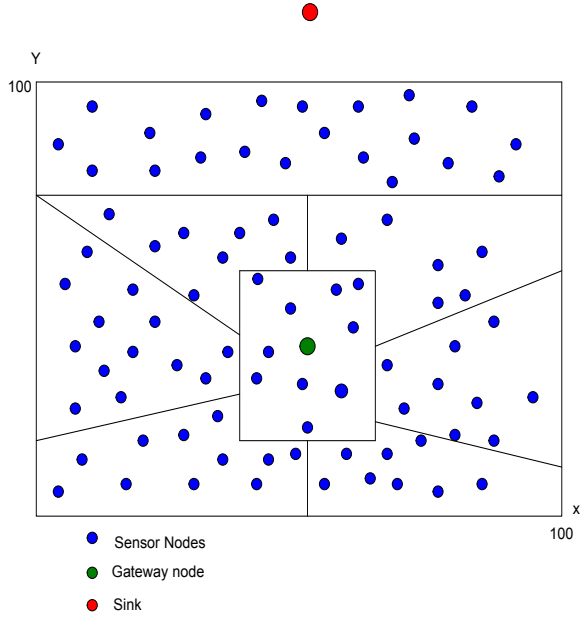}
\caption{Network Model}
\end{center}
\end{figure}

We use first order radio model as used in~\cite{heinzelman2000energy} and~\cite{heinzelman2002application}. This model represents the energy dissipation of sensor nodes for transmitting, receiving and aggregating data. The transmitter dissipates more energy then receiver as it requires more energy for the transmitter electronics and amplifier. On the other hand, in receiver, only electronic circuit dissipate energy, as shown in fig 2.\\
\begin{figure}[ht]
\begin{center}
\includegraphics[scale=0.6]{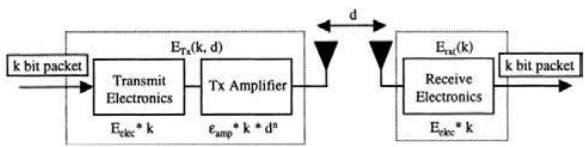}
\caption{Radio Model}
\end{center}
\end{figure}

The energy required to transmit a data packet of ${k}$ bits to a distance ${d}$ and to receive a data packet of ${k}$ bits, is given as:\\

\begin{equation*} %
E_{Tx}(k,d) = E_{Tx-elec}(k) + E_{Tx-amp}(k,d)
\end{equation*}
\begin{equation}\label{1}
E_{Tx}(k,d) = E_{elec}\times k + E_{amp}\times k\times d^2
\end{equation}

\begin{equation*}
E_{Rx}(k) = E_{Rx-elec}(k)
E_{Rx}(k) = E_{elec} \times k
\end{equation*}
\begin{equation}\label{2}
E_{Rx}(k) = E_{elec} \times k
\end{equation}

\section{The M-GEAR Protocol}

In this section, we present detail of our proposed protocol. Sensor nodes have too much sensed data for BS to process. Therefore, an automatic method of combining or Aggregating the data into a small set of momentous information is required~\cite{hall2004mathematical}~\cite{klein1993sensor}. The process of data aggregation also termed as data fusion.
In order to improve network lifetime and throughput, we deploy a gateway node at the centre of the network field. Function of gateway node is to collect data from CHs and from nodes near gateway, aggregation and sending to BS. Our results ensure that network lifetime and energy consumption improved with the expense of adding gateway node. We add rechargeable gateway node because it is on ground fact that the recharging of gateway node is much cheaper than the price of sensor node.\\

\subsection{Initial Phase}
 In M-GEAR, we use homogenous sensor nodes that are dispersed randomly in network area. The BS broadcast a HELLO packet. In response, the sensor nodes forward their location to BS. The BS calculates the distance of each node and save all information of the sensor nodes into the node data table. The node data table consists of distinctive node ID, residual energy of node, location of node and its distance to the BS and gateway node.\\

\subsection{Setup Phase}

In this section, we divide the network field into logical regions based on the location of the node in the network. BS divide the nodes into four different logical regions. Nodes in region-one use direct communication and transmit their data directly to BS as the distance of these nodes from BS is very short. Similarly nodes near gateway form region-two and send their data directly to gateway which aggregates data and forward to BS. These two regions are referred to as non clustered regions.  All the nodes away from the gateway node and BS are divided into two equal half regions. We call them clustered regions. Sensor nodes in each clustered region organize themselves into small groups known as clusters.\\

\subsection{CH Selection}
Initially BS divides the network into regions. CHs are elected in each region separately. Let r$_{i}$  represent the number of rounds to be a CH for the node S $_{i}$. Each node elect itself as a CH once every r $_{i}$ = 1/p rounds.
At the start of first round all nodes in both regions has equal energy level and has equal chance to become CH. After that CH is selected on the basis of the remaining energy of sensor node and with a probability p alike LEACH. in each round, it is required to have n x p CHs. A node can become CH only once in an epoch and the nodes not elected as CH in the current round feel right to the set C.  The probability of a node to (belongs to set C) elect as CH increases in each round. It is required to uphold balanced number of CHs.
At the start of each round, a node S$_{i}$ belongs to set C autonomously choose a random number between 0 to 1. If the generated random number for node S$_{i}$  is less than a predefined threshold T(s) value then the node becomes CH in the current round.\\
The threshold value can be found as:\\
\begin{equation}
T(S)=\begin{cases}
\frac{p}{1-p \times(rmod(1/p))} & \text{ if } s \in C \\
0 & \text otherwise
\end{cases}
\end{equation}
where P = the desired percentage of CHs and r = the current round, C = set of nodes not elected as CH in current round.
After electing CHs in each region, CHs inform their role to neighbor nodes. CHs broadcast a control packet using a CSMA MAC protocol. Upon received control packet from CH, each node transmits acknowledge packet. Node who find nearest CH, becomes member of that CH.\\

\subsection{Scheduling}
When all the sensor nodes are structured into clusters, each CH creates TDMA based time slots for its member nodes. All the associated nodes transmit their sensed data to CH in its own scheduled time slot. Otherwise nodes switch to idle mode. Nodes turn on their transmitters at time of transmission. Hence, energy dissipation of individual sensor node decreases.\\

\subsection{Steady-State Phase}
In steady state phase, all sensor nodes transmit their sensed data to CH. The CH collects data from member nodes, aggregates and forwards to gateway node. Gateway node receives data from CHs, aggregates and forwards to BS.\\

\section{Performance Evaluation}
We assess the performance of our proposed protocol and compare it with existing protocol in WSN, known as LEACH.

\subsection{Simulation Setting}

In order to appraise the performance of our proposed protocol, we simulated our protocol using MATLAB. We consider a wireless sensor network with 100 nodes distributed randomly in 100m X 100m field. A gateway node is deployed at the centre of the sensing field. The BS is located faraway from the sensing field. Both gateway node and BS are stationary after deployment. We consider packet size of 4000 bits. We compare our protocol with LEACH protocol. To assess performance of our protocol with LEACH, we ignore the effects caused by signal collision and interference in the wireless channel. Table 1 presents the radio parameters.

\subsection{Performance Parameters}
In this subsection, we present performance metrics. In this work, we evaluated three performance parameters given below.
\subsubsection{Network lifetime}
It is the time interval from the start of the network operation till the last node die.\\
\subsubsection{Throughput}
To evaluate the performance of throughput, the numbers of packets received by BS are compared with the number of packets sent by the nodes in each round.\\
\subsubsection{Residual Energy}
The residual battery energy of network is considered in order to analyze the energy consumption of nodes in each round. Residual energy ensures graceful degradation of network life.\\

\subsection{Simulation Results and Analysis}
In this subsection, we show the simulation results. We run extensive simulations and compare our results with LEACH. Next subsections give detail of each metric.
\subsubsection{Network Lifetime}
In fig 3, we show the results of the network lifetime. Nodes are considered dead after consuming 0.5 joule energy. M-GEAR protocol obtains the longest network lifetime. This is because the energy consumption is well distributed among nodes. Network is divided into logical regions and two of them are further sub divided into clusters. M-GEAR topology balance energy consumption among sensor nodes. On the other hand, in LEACH, nodes die quickly as stability period of network ends. It is not evident that predestined CHs in LEACH are distributed uniformly throughout the network field. Therefore, there is a possibility that the selected CHs will be concentrated in one region of the network. Hence, some nodes will not have any CHs in their environs. Fig 3 shows interval plot of network lifetime with 99\% confidence interval. we note that, the results of M-GEAR protocol are statically different and perform well.

\begin{figure}[!h]
\begin{center}
\includegraphics[scale=0.5]{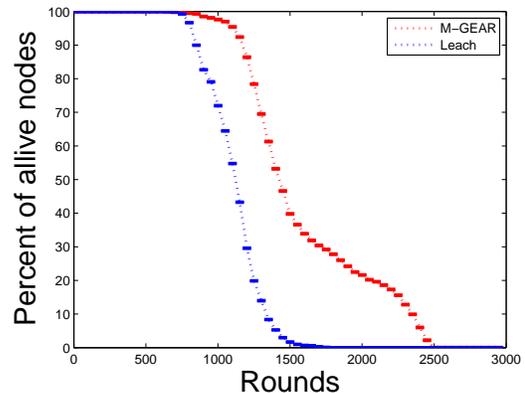}
\caption{Interval plot- Analysis of network lifetime}
\end{center}
\end{figure}

\subsubsection{Throughput}
Average packets sent to BS are assessed through extensive simulations. Simulation results of M-GEAR protocol illustrate increased throughput. Interval plots of M-GEAR and LEACH in fig 4 clearly depicts performance of both protocols. To calculate throughput, we assume that CHs can communicate freely with gateway node. Simulation results show an increase throughput of 5 times then LEACH.
Sensor nodes near gateway send their data directly to gateway; similarly nodes near BS transmit data directly to BS. Sensor nodes in both regions consume less transmission energy therefore, nodes stay alive for longer period. More alive nodes contribute to transmit more packets to BS.

\begin{figure}[!h]
\begin{center}
\includegraphics[scale=0.5]{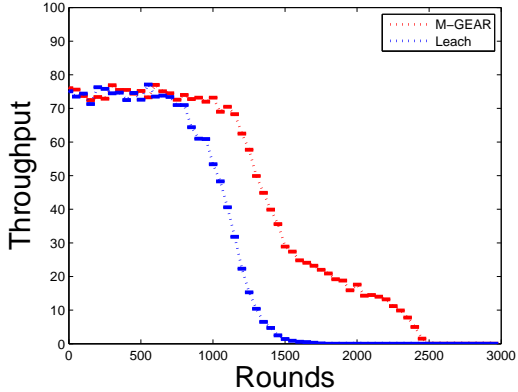}
\caption{Interval plot- Analysis of Throughput}
\end{center}
\end{figure}

\subsubsection{Residual Energy}
Fig 5 shows average residual energy of network per round. We assume that a node has 0.5 joule energy. The total energy of 100 node network is 50 joule. M-GEAR protocol yields minimum energy consumption than LEACH. Fig 5 clearly depicts that our protocol outperforms LEACH routing protocol in terms of energy consumption per round. Deployment of gateway node at the centre and high probability of CHs in all regions ensures minimum energy consumption.

\begin{figure}[!h]
\begin{center}
\includegraphics[scale=0.5]{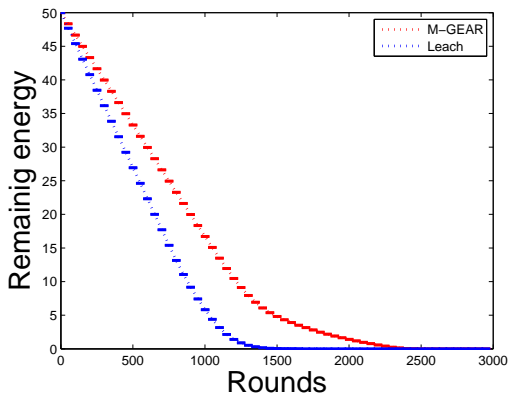}
\caption{Interval plot- Analysis of remaining energy}
\end{center}
\end{figure}

\section{Conclusion and future work}
We describe an energy-efficient multi-hop routing protocol using gateway node to minimize energy consumption of sensor network. In this work, we divide the network into logical regions. Each region use different communication hierarchy. Two regions use direct communication topology and two regions are further sub-divided into clusters and use multi-hop communication hierarchy.  Each node in a region elects itself as a CH independent of other region. This technique \begin{table}
  \centering
  \begin{tabular}{|l|c|r|}
     \hline
     \textbf{Parameter}   &     \textbf{Value}\\
\hline

          Eo      & 0.5j \\
          \hline
      E\emph{elec}  &   5nJ/bit \\
      \hline
      E\emph{fs}   &   10pJ/bit/m2\\
      \hline
      E\emph{mp}   &     0.0013 pJ/bit/m4 \\
      \hline
      Eda &            5pJ/ bit\\
      \hline
      Message size & 4000 bits \\
  \hline

 \end{tabular}%
\caption{Network parameter}
 \end{table}
 encourges better distribution of CHs in the network. Simulation results shows that our proposed protocol performs well compared to LEACH. In this work, we study the three performance metrics: Network lifetime , Residual energy and throughput. In future, we will study ETX link metrics and we will implement this metric in our scheme as implemented and demonstrated in~\cite{javaid2009performance}~\cite{dridi2010performance}~\cite{dridi2009ieee}.

\end{document}